# Large-area van der Waals epitaxy and magnetic characterization of $Fe_3GeTe_2$ films on graphene

J. Marcelo J. Lopes,[1,*] Dietmar Czubak,[1] Eugenio Zallo,[1,†] Adriana I. Figueroa,[2] Charles Guillemard,[3] Manuel Valvidares,[3] Juan Rubio Zuazo,[4,5] Jesús López-Sanchéz,[4,5] Sergio O. Valenzuela,[2,6] Michael Hanke,[1] Manfred Ramsteiner[1,*]

1- Paul-Drude-Institut für Festkörperelektronik, Leibniz-Institut im Forschungsverbund Berlin e.V., Hausvogteiplatz 5-7, 10117 Berlin, Germany
2- Catalan Institute of Nanoscience and Nanotechnology (ICN2), CSIC and BIST, Campus UAB, Bellaterra, 08193 Barcelona, Spain
3- ALBA Synchrotron Light Source, Barcelona 08290, Spain
4- Spanish CRG BM25-SpLine at The ESRF – The European Synchrotron, 38000 Grenoble, France
5- Instituto de Ciencia de Materiales de Madrid (ICMM), CSIC, 28049 Madrid, Spain
6- Institució Catalana de Recerca i Estudis Avançats (ICREA), Barcelona 08010, Spain

**Abstract**

Scalable fabrication of magnetic 2D materials and heterostructures constitutes a crucial step for scaling down current spintronic devices and the development of novel spintronic applications. Here, we report on van der Waals (vdW) epitaxy of the layered magnetic metal $Fe_3GeTe_2$ – a 2D crystal with highly tunable properties and a high prospect for room temperature ferromagnetism – directly on graphene by employing molecular beam epitaxy. Morphological and structural characterization confirmed the realization of large-area, continuous $Fe_3GeTe_2$/graphene heterostructure films with stable interfaces and good crystalline quality. Furthermore, magneto-transport and X-ray magnetic circular dichroism investigations confirmed a robust out-of-plane ferromagnetism in the layers, comparable to state-of-the-art exfoliated flakes from bulk crystals. These results are highly relevant for further research on wafer-scale growth of vdW heterostructures combining $Fe_3GeTe_2$ with other layered crystals such as transition metal dichalcogenides for the realization of multifunctional, atomically thin devices.

---

[*] Corresponding Authors: lopes@pdi-berlin.de; ramsteiner@pdi-berlin.de
[†] Current affiliation: Walter Schottky Institut and Physik Department, Technische Universität München, Am Coulombwall 4, 85748 Garching, Germany

## 1- Introduction

Layered magnetic materials such as $Cr_2Ge_2Te_6$ [Gon17], $CrI_3$ [Hua17], $CrTe_2$ [Pur20], and $Fe_3GeTe_2$ [Den18] are considered to be promising building blocks for the development of ultra-compact spintronic devices with faster response and low-power dissipation. Moreover, their study is expected to open new perspectives on a more versatile modulation of magnetic properties, beyond what can be achieved in traditional 3D magnetic thin films [Zut04,Gib19,Gon19]. As an example, combining these layered magnets with other 2D crystals such as graphene, transition metal dichalcogenides (TMDCs), or metal monochalcogenides to form van der Waals (vdW) heterostructures offers great potential to tailor magnetism via proximity-induced phenomena [Chen19,Dol20,Kar20,Gon19].

2D magnets and related vdW heterostructures promise a large impact in future applications if they can sustain magnetic order up to room temperature [Val19]. Within the current catalogue of available materials, one of the most promising candidates to fulfill this requirement is $Fe_3GeTe_2$ (FGT) [Den18,Wan20]. FGT is a ferromagnetic conductor with a robust out-of-plane anisotropy [Fei18] and, most remarkably, highly tunable properties. It has been shown that the Curie temperature ($T_C$) in few-layer thick FGT can be boosted (from around 100 K) to room temperature, or even beyond that (up to 400 K), via electrostatic doping [Den18], or when grown onto a topological insulator [Wan20]. Nearly room temperature ferromagnetism (FM) could also be achieved by modulating the composition during growth to realize the Fe-rich phases $Fe_4GeTe_2$ and $Fe_{5-x}GeTe_2$ [May19,Seo20]. Besides these possibilities, strain engineering and substitutional doping have also been identified as effective routes to exert control over the magnetic properties of FGT [Zhu16,Hu20,Tia20]. Finally, although a native surface oxide forms upon air exposure [Kim19], degradation in FGT is not as severe as in chromium halides [Shc18], which should facilitate its realistic exploration in devices.

As for other magnetic 2D materials, experimental research on FGT has mainly concentrated on micrometer-sized flakes exfoliated from bulk crystals. This type of sample allows for exploring fundamental properties as well as device concepts based on vdW heterostacks [Lin20]. However, it is not compatible with standard device fabrication, which usually requires a wafer-scale, uniform form of material. Recent progress has been made in this direction, with a few studies reporting on molecular beam epitaxy (MBE) of FGT films on $Al_2O_3(0001)$ [Liu17,Wan20], GaAs(111) [Liu17], and Ge (111) [Roe20]. MBE growth of FGT/MnTe [Liu17] and FGT/$Bi_2Te_3$ [Wan20] heterostructures, both on $Al_2O_3(0001)$ wafers, has also been demonstrated. In general, the collection of data contained in these studies suggests - as usual in conventional epitaxy - that lattice-matching between FGT and its growth template is crucial to obtain large-area films with good crystalline quality. However, such requirement also reduces the number of choices in terms of suitable materials that can be employed as substrate for FGT, or combined with it to realize magnetic superlattices. Interestingly, this seems to hold true even for $Bi_2Te_3$-buffered $Al_2O_3(0001)$ templates [Wan20], on which FGT growth should in principle proceed via vdW epitaxy [Wal17]. The detection of strain in the FGT film has been associated with the lattice mismatch between both materials [the ($a = b$) in-plane lattice constant of $Bi_2Te_3$ and FGT is 4.38 Å and 3.99 Å, respectively)] [Hos19,Dei06], indicating that interlayer interactions beyond vdW-bonding take place at the $FGT/Bi_2Te_3$ interface. In spite of this observation, the weak nature of vdW interactions is generally believed to alleviate the need for lattice matching conditions [Kom92], so that via vdW epitaxy of FGT on top of 2D crystals such as graphene, hexagonal BN (h-BN), and TMDCs should enable the transfer-free, scalable realization of 2D heterostructures exhibiting atomically smooth vdW interfaces. Nevertheless, the intrinsically weak bonding between 2D materials also creates difficulties namely uncontrolled nucleation on morphological defects (e.g., wrinkles) [Hei18,Lin14], which can in turn result in non-uniform growth of low-quality material. Hence, exploiting epitaxial growth of FGT on other 2D crystals is greatly demanded for the development of atomically thin magnetic vdW materials to be implemented into future applications.

In this work, we assess the feasibility of growing FGT thin films directly on graphene surfaces by using MBE. We employed different characterization tools to investigate the structure and magnetic properties of the FGT/graphene heterostructures. Large-scale, homogeneous FGT films exhibiting a smooth surface morphology could be realized on graphene with a preferential epitaxial orientation and low in-plane mosaicity. Additionally, no structural damage or modification took place in graphene due to FGT growth, indicating the formation of a stable vdW interface between the materials. Magnetism in the samples were probed macro- and microscopically, revealing a ferromagnetic behavior for the FGT films that is similar to state-of-the art exfoliated flakes as well as thin films grown on 3D substrates. Overall, our results demonstrate that vdW epitaxy of FGT on graphene is a powerful approach for the wafer-scale fabrication of magnetic vdW heterostructures combining dissimilar 2D crystals for spintronics.

## 2- Experimental Section

FGT films were synthesized by MBE (base pressure around $5 \times 10^{-11}$ mbar) using elemental Fe, Ge, and Te evaporated from Knudsen cells. For all growth experiments the flux ratio of Fe, Ge, and Te was kept at ~ 2:1:20. They were obtained by measuring the respective beam equivalent pressures employing a pressure gauge. Continuous films with thicknesses around 10 and 20 nm (i.e. from 12 to 25 quintuple layers (QL); 1 QL ~ 0.8 nm formed by sequential Te/Fe/FeGe/Fe/Te slabs) were obtained using growth times of 1 and 2 hours, respectively. As substrates, 1 cm × 1cm large epitaxial graphene on-axis semi-insulating 4H-SiC(0001) were utilized. They were fabricated via SiC surface graphitization at high temperatures in an Ar atmosphere, following well-established synthesis protocols [Emt09,Oli11]. FGT films were also synthesized on 1 cm × 1cm $Al_2O_3$ (0001) pieces using the same conditions for comparison. The graphene/SiC(0001) substrates were outgassed at 450 °C (550 °C for sapphire) for at least one hour and then cooled to 300 °C for FGT growth. All substrates were coated with ~ 1 μm of Ti on the backside via electron beam evaporation to allow non-contact heating by radiation. *In-situ* growth monitoring was performed by reflection high energy electron diffraction (RHEED). Finally, in some cases the FGT films were capped *in-situ* with a ~ 5 nm thick amorphous Te layer deposited after sample cooling to room temperature. This procedure was adopted in order to reduce FGT surface oxidation upon prolonged air exposure [Roe20], which could hinder the successful use of the surface-sensitive techniques X-ray absorption spectroscopy (XAS), X-ray magnetic circular dichroism (XMCD), and grazing incidence X-ray diffraction (GID) (see below).

Atomic force microscopy (AFM) in tapping mode was employed to probe the surface morphology of the FGT films. Raman spectroscopy in back scattering configuration was performed with the aim of obtaining information about the structure of FGT and graphene. Raman measurements were also carried out on commercially available FGT bulk crystals (HQ graphene) for comparison. The Raman spectra were excited at a wavelength of 473 nm with the laser beam focused onto the sample surface to a 1 μm diameter spot. In order to probe the azimuthal dependence of various in-plane lattice parameters, we have used synchrotron-based GID. This experiment was performed at the SpLine beamline BM25 at the European Synchrotron Radiation Facility (ESRF) in Grenoble. An incidence angle of the illuminating X-rays of 0.2° sufficiently suppresses strong scattering by the substrate, and thus makes this method highly surface sensitive. A primary X-ray photon energy of 17 keV ($\lambda$ = 0.7293 Å) enables the inspection of a comparatively large area in reciprocal space, which is important to probe also multiple reflections of the same lattice plane family.

Magneto-transport characterization was performed within the 4-300 K temperature range and with magnetic field up to 8 T, using a large-area van der Pauw (vdP) geometry with Al contact wires bonded at the edges of the 1 cm × 1 cm FGT film. XAS and XMCD measurements were employed to investigate the magnetic properties microscopically [Figu14]. XAS spectra at the Fe $L_3$ and $L_2$ edges were recorded in the beam line BL-29 (BOREAS) at the ALBA synchrotron (Spain), which provides a ultra-high-vacuum sample environment with a base temperature of ~ 3 K and magnetic field up to 6 T. Measurements used total-electron-yield detection, where the drain current was measured from the sample to the ground. The magnetic field was applied along the X-ray beam at normal incidence relative to the sample plane. XMCD was obtained by subtracting XAS spectra with the photon helicity vector antiparallel and parallel to the magnetic field.

## 3- Results and Discussion

AFM analyses revealed the formation of continuous FGT films on graphene/SiC(0001) without strong thickness inhomogeneity due to island formation or localized out-of-plane-growth. Figure 1a shows a typical AFM height image of a ~ 10 nm thick FGT film which demonstrates its coverage over the micrometer-large graphene/SiC surface terraces. Very similar results were obtained for 20 nm thick FGT. Discontinuities (holes) in the film are observed mostly close to step edges (indicated by red arrows), resulting in relatively small areas with exposed graphene (see Figure 1b). The absence of FGT in these regions is confirmed by phase-contrast imaging (not shown), as well as by the depth profile shown in Figure 1b (inset).

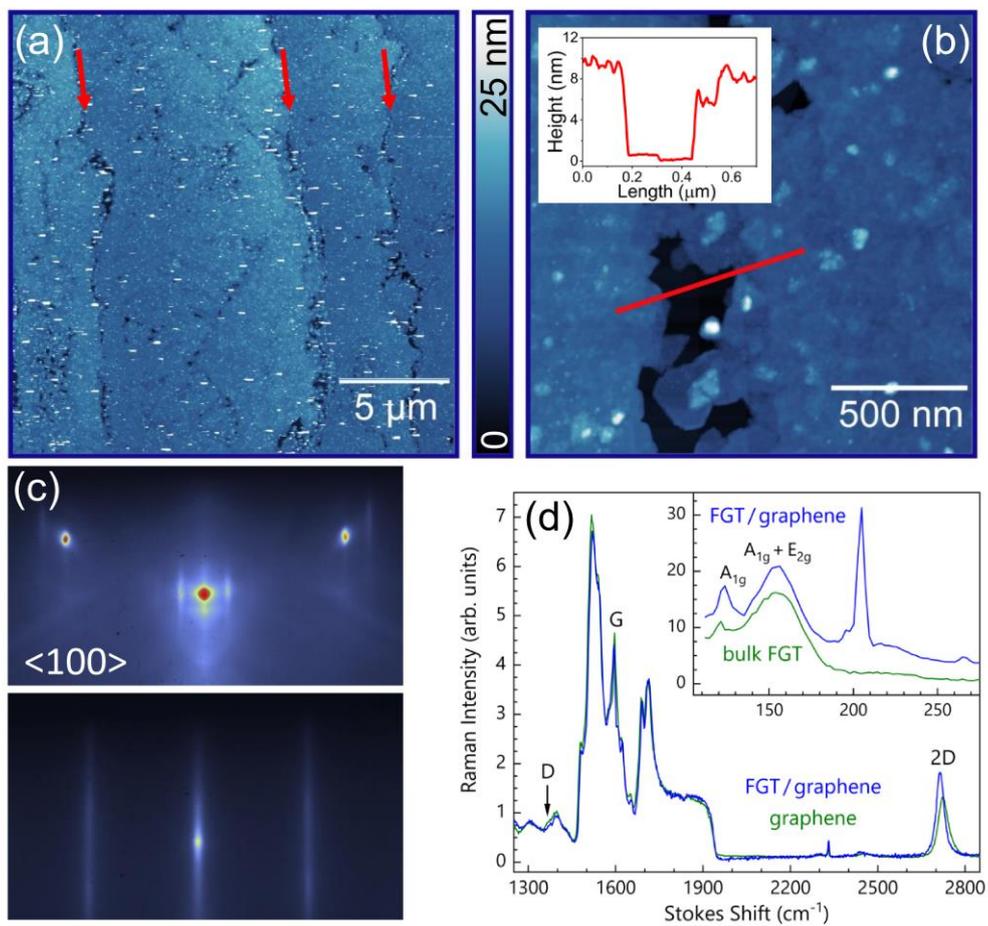

**Figure 1.** (a) AFM height image of a ~ 10 nm thick FGT grown on graphene/SiC(0001). The red arrows indicate the position of surface steps in graphene/SiC(0001). (b) AFM image of a surface area close to a step edge in graphene/SiC(0001) where a discontinuity in FGT is evident. The inset shows the profile corresponding to the red line in the main figure. (c) In-situ RHEED patterns taken perpendicular to the SiC<100> direction, before (upper panel) and after FGT growth (lower panel). (d) Raman spectra showing the graphene and SiC components for samples with (blue spectrum) and without (green spectrum) a ~ 20 nm thick FGT film on top. The inset depicts the $A_{1g}$ and $A_{1g} + E_{2g}$ Raman components associated with the FGT film (blue spectrum). For comparison, the Raman spectrum collected from a bulk FGT crystal is also plotted (green spectrum).

Interestingly, a small step of ~ 0.4 nm is visible in the valley region of the line profile, which is known to form at the border between *n* and *n+1* graphene layers as a result of the SiC decomposition process during graphene formation [Ara17]. The appearance of such morphological irregularities in the FGT film is probably related to an inability of FGT to grow - under the employed synthesis conditions - over the bi- to few-layer thick graphene ribbons and patches located on step edge regions [Emt09,Oli11,Boe15,Ara17]. They exhibit a much lower chemical reactivity in comparison to that of monolayer graphene covering surface terraces. This variation is known to affect the overgrowth of 2D materials [Hei20]. Further optimization of FGT growth parameters and graphene substrate preparation are anticipated to mitigate this problem. The root-mean-square (RMS) roughness of the FGT film was measured to be ~ 1 nm in 1 µm × 1 µm on a surface terrace, which is higher than that measured for an area of same size in graphene (RMS ~ 0.2 nm). Note that the AFM images depicted here were acquired in non-capped FGT films exposed to air for several days. Thus, natural oxidation of the topmost few layers of the FGT film [Kim19], as well as water and other surface contaminants, are expected to affect the surface roughness.

Figure 1c exhibits a typical RHEED pattern obtained during FGT growth (bottom panel), where the narrow streaks indicate epitaxial growth of a 2D film with smooth surface and good structural quality. In addition, the shift in their location with respect to the pattern obtained for the graphene/SiC template (upper panel) reveals the evolution to a larger in-plane lattice constant, as expected for FGT growth (see GID later).

Raman spectroscopy (see Figure 1d) confirmed that the graphene layer did not suffer structural changes due to FGT synthesis. Besides the SiC background signal, the graphene-related 2D and G peaks remain mostly unaltered. The small variation in the position and width of the 2D peak is probably associated with excitation of surface areas containing a different number of graphene layers, e.g., close to step edge regions [Emt09,Oli11,Hei20]. Importantly, the intensity of the defect-related D peak does not increase after growth. This confirms the structural integrity of graphene and indicates that no rehybridization takes place in graphene, e.g., due to the formation of covalent bonds with FGT. Finally, the Raman measurements also served to confirm the formation of FGT. The inset in Figure 1d shows the spectral region that is relevant for this material. The two components appearing at ~ 120 and ~ 155 $cm^{-1}$ are associated with the $A_{1g}$ and the sum of the $A_{1g}$ + $E_{2g}$ phonon modes in FGT [Du19]. The $A_{1g}$ mode for the FGT is slightly blue-shifted with respect to that of bulk FGT. This might be associated with the strain in the FGT film (see GID results below), or due to different degrees of degradation of the materials due to surface oxidation and/or laser exposure [Du19]. Investigating these issues in detail requires future dedicated experiments. The strong peak located around 205 $cm^{-1}$ originates from the SiC substrate [Bur98].

Further information on the structure of FGT/graphene heterostacks was obtained by GID. Figure 2a shows an in-plane reciprocal space map of a Te-capped, ~ 10 nm thick FGT film, in which the color-coded scattered intensity in reciprocal space is plotted as a function of reciprocal lattice units (rlu), referring to the hexagonal lattice of the SiC(0001) substrate. The corresponding *absolute* length (in units of Å$^{-1}$) of the scattering vector is additionally provided along the H$^{hex}$ axis. There are highly localized substrate reflections, namely the SiC($2\bar{1}$.0) reflection and its higher orders. Over imposed to the map itself there is a line scan crossing the SiC($2\bar{1}$.0) reflection. Note that we use the four-component vector notation for hexagonal symmetry, *i.e.*, (hklm) = (hk.m) = (hk-(h+k)m). As a guide for the eye we have drawn different radii corresponding to net planes distances from graphene, G(10.0) and G(11.0), and FGT, *i.e.*, the FGT(n0.0) with n=[2..4] and FGT(nn.0) with [n=1..3]. From the radial scan one can extract quantitative information on the various in-plane lattice spacing values. Based on the three most intense contributions from FGT [i.e., (11.0), (30.0) and (22.0)] one can deduce an average in-plane lattice parameter for FGT of 4.011 Å, which indicates a slight tensile strain of about 0.5% (taking as reference value d$_{FGT}$ = 3.991 Å for bulk crystals [Dei06]). This tensile strain is much smaller than what has been reported for FGT grown on Bi$_2$Te$_3$ (1.7 % tensile strain) [Dei06], in agreement with the anticipated weak interaction between FGT and graphene. We speculate that the observed strain might be related to structural and morphological deformations in the FGT layers, such as bending taking place around surface irregularities (graphene wrinkles) and the SiC surface steps.

Following the intensity distribution along the FGT arcs [e.g., the FGT(22.0)], one recognizes a pronounced preferential azimuthal orientation of the FGT layer in a way that the FGT[11.0] net planes are parallel to SiC[11.0]. Two of those angular profiles crossing the FGT(11.0) and (22.0) reflection are plotted in Figure 2b. They prove a full width at half maximum of about 4.7°, indicating a FGT lattice well-aligned with respect to the substrate. Additionally, there are weaker maxima 30° off revealing a smaller fraction of the FGT layer which appears twinned to the major variant. Complementary to that, arcs corresponding to FGT(n0.0) (see Figure 2a) have their maxima at K = 0, which fully confirms the previous statement. Moreover, there are additional low-intensity maxima symmetric to the FGT(22.0) at 0° and ±60°, which are about 10° off the ($2\bar{1}$.0) direction. These highly symmetric sub-features are most probably due to a coincidence lattice between the FGT layer and the underlying graphene and/or SiC substrate [Boe15]. The very same coincidence lattice might be responsible for the additional arcs next to the SiC($2\bar{1}$.0) and ($4\bar{2}$.0), which contain pronounced maxima azimuthally 20° off the [$2\bar{1}$.0] direction, or in other words: they are 10° off the [10.0] direction.

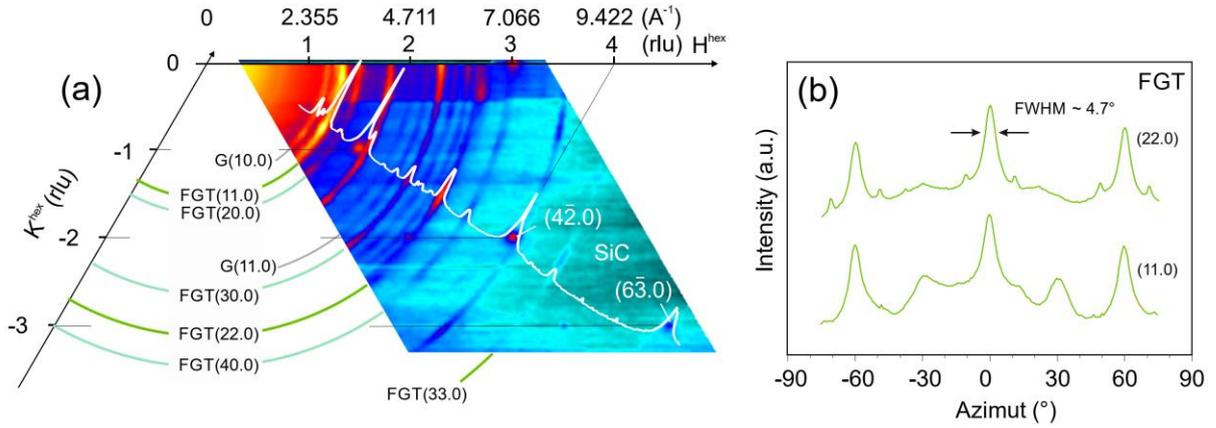

**Figure 2.** (a) GID in-plane reciprocal space map of a Te-capped 10 nm thick FGT on graphene/SiC(0001), reflecting the azimuthal dependence of various in-plane lattice parameters. The white line shows a linear scan along the SiC[$2\bar{1}.0$] direction. (b) Two azimuthal intensity profiles intersecting the maxima caused by the FGT lattice.

The magnetic and electrical characteristics of FGT/graphene heterostacks were investigated by transverse resistance ($R_{XY}$) measurements. In ferromagnetic materials, $R_{XY}$ under the application of an out-of-plane magnetic field H is given by the superposition of the ordinary (OHE) and anomalous (AHE) Hall effects and can be expressed as

$R_{XY} = R_{OHE} + R_{AHE}$   with $R_{OHE} = \mu_0 R_O H$, and $R_{AHE} = \mu_0 R_A M$ (1)

where $\mu_0$ is the vacuum permeability and M the magnetization. $R_O$ and $R_A$ are the ordinary and anomalous Hall coefficients, respectively [Che08]. Whereas the OHE ($R_{OHE}$) contains information about the electrical characteristics, the AHE ($R_{AHE}$) can be utilized to investigate the magnetic order and the magnetization reversal in ferromagnetic films. Figure 3a displays $R_{XY}$ for a 20 nm thick FGT on graphene measured at different temperatures during subsequent downward and upward sweeps of an external magnetic field. At low temperature (5 K), the AHE is clearly detected as a square shape hysteresis loop superimposed on an OHE contribution with its linear dependence on the external field (see Eq. 1) [Tan18,Roe20]. The occurrence of the AHE reflects the ferromagnetic order in the FGT film, with the observed remanence providing evidence for a strong perpendicular magnetic anisotropy. In agreement with Ref. [Roe20], both the coercive fields $H_C$ as well as the saturation resistance $\Delta R_{AHE}$ decrease with increasing temperature. Thereby, $\Delta R_{AHE}$ is defined as the value of $R_{AHE}$ (see Eq. 1) at magnetization saturation. The temperature dependencies of $H_C$ and $\Delta R_{AHE}$ are shown in Figure 3b. The extrapolation of both quantities to zero indicates in average a Curie temperature of about 220 K that agrees well with what has been reported for FGT nanoflakes [Tan18].

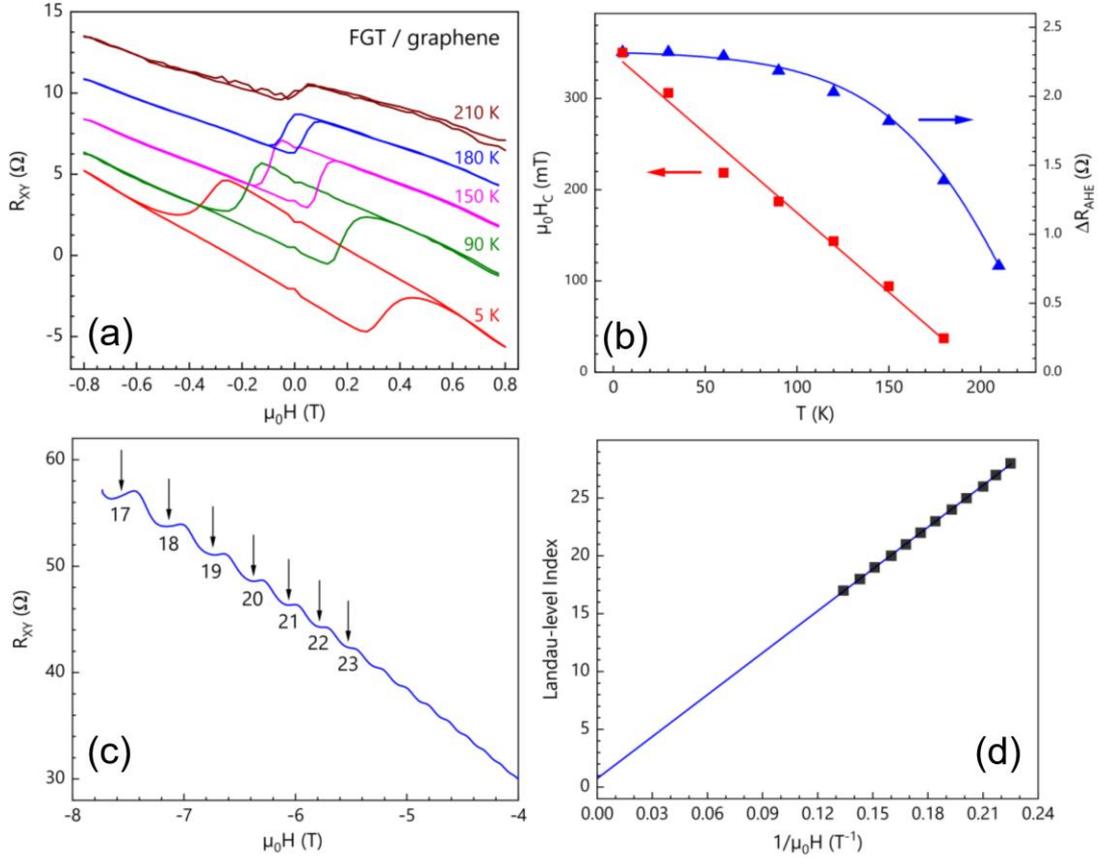

**Figure 3.** (a) Transverse resistance $R_{XY}$ of a 1 cm x 1 cm FGT/graphene vdP structure during downward and upward magnetic field sweeps at different temperatures. The $R_{XY}$ curves for temperatures above 5 K are successively upshifted by 2.5 Ω for clarity. (b) Coercive field $H_C$ and saturation resistance $\Delta R_{AHE}$ extracted from the AHE contribution as a function of temperature. The solid lines are guides to the eye. (c) Transverse resistance $R_{XY}$ of the same FGT/graphene vdP structure as a function of magnetic field in the range from -4 to -8 T. (d) Landau-level index of the QHE plateaus in (c) as a function of the inverse magnetic field. The solid line represents the result of a linear fitting.

From the slope of the OHE contribution to the Hall curves shown in Figure 3a, a low-temperature carrier density of $6.8 \times 10^{13}$ cm$^{-2}$ is extracted which, in principle, can be ascribed to highly n-type doped graphene on SiC(0001). in contrast, FGT films are expected to exhibit a hole-like transport character with an opposite $R_{OHE}$ slope and have a much larger carrier density (on the order of $2 \times 10^{15}$ cm$^{-2}$), as determined for films on sapphire substrates (see *Supplementary Materials*), which is in agreement with Ref. [Liu17]. The sheet resistivity of the FGT/graphene heterostructure (150 Ω/sq) is found to be almost one order of magnitude smaller than the one commonly observed for our epitaxial graphene films (~ 2 kΩ/sq). These findings are explained consistently by considering parallel conduction through two transport channels for the analysis of the OHE [Far11,Har11]. In this case, the sign as well as the absolute value of the $R_{OHE}$ slope depend on the relative conductivities of the FGT and graphene films.

Therefore in general they do not reflect the carrier type and density of only one of the two transport channels [Har11].

Figure 3c reveals that the transverse resistance $R_{XY}$ actually exhibits plateaus which can be explained by the quantum Hall effect (QHE) in graphene or precursors of the QHE occurring at magnetic fields with magnitudes above 4 T [Job10]. Please note that the $R_{XY}$ data shown in Figure 3c contain a certain contribution of the magnetic-field dependent longitudinal resistance due the utilized vdP geometry. The positions of the plateaus can be associated with Landau-level indices and exhibit the expected linear dependence on the inverse field $(\mu_0 H)^{-1}$ [Job10,Zha05], as can be seen in Figure 3d. The occurrence of the QHE plateaus underlines the high quality and carrier mobility of the graphene film underneath FGT over the large-area vdP structure.

XMCD/XAS results obtained for a Te-capped, 20 nm thick FGT film on graphene/SiC(0001) are shown in Figure 4. The magnetic hysteresis loop shown in Figure 4a was recorded at base temperature of 3 K with the beam energy fixed at the maximum of the Fe $L_3$ XMCD signal (706.4 eV) (see inset Figure 4b). The signal is normalized to an energy point at the pre-edge region (703 eV) [Goe00], measured on the fly with a magnetic field ramp of 1.5 T/min at high $B$ (from ± 6 to ± 2 T) and of 1.0 T/min at low $B$ (from ± 2 to ∓ 2 T). In accordance with the magnetotransport results, the square shape of the loop reveals the strong out-of-plane ferromagnetic character of the Fe atoms in the FGT film.

The temperature dependence of the maximum of the Fe $L_3$ XMCD signal in saturation (6 T) and remanence (0 T) is plotted in Figure 4b. The remanent magnetization extrapolates to zero at about ∼ 220 K, which agrees closely with the $T_C$ found by transport measurements. The 6-T XMCD signal persists at room temperature, demonstrating a strong paramagnetic response even though long range ferromagnetic order has already vanished. Application of the magneto-optical sum rules for XAS and XMCD [Tho92,Car93,Che95] allows us to extract the orbital, spin and total moments ($m_{orb}$, $m_{spin}$ and $m_{Total}$) of Fe 3$d$ electrons in the ground state of FGT. The number of 3$d$ holes was set to 4, as reported for the Fe $3d^6$ configuration in Fe$_3$GeTe$_2$ [Zhu16], and the value of the magnetic dipole operator term was neglected. From the experimental Fe $L_{2,3}$ edges in the spectra at 3 K and 6 T we obtain $m_{orb}$ = 0.047 ± 0.004 $\mu_B$/Fe; $m_{spin}$ = 1.13 ± 0.09 $\mu_B$/Fe and $m_{Total}$=1.18 ± 0.10 $\mu_B$/Fe. The orbital-to-spin moment ratio is 0.042 ± 0.008, which is comparable to that of elemental Fe (0.043) [Che95], somewhat larger than what is reported for Fe$_3$GeTe$_2$ (0.03) [Zhu16], but lower than that of Fe$_5$GeTe$_2$ (0.056) [Yam21].

In summary, vdW epitaxy of FGT magnetic films on graphene was investigated by employing molecular beam epitaxy. The realization of large-area FGT/graphene heterostructure films with high structural and interface quality was demonstrated via a thorough structural characterization, as well as magneto-transport and XMCD investigations, which revealed a robust perpendicular magnetic anisotropy in FGT and the observation of plateaus associated with the QHE effect in epitaxial graphene. These results are highly relevant for further research on wafer-scale growth of atomically thin, all-epitaxial vdW heterostructures with multifunctional properties. By combining $Fe_3GeTe_2$ with other selected 2D crystals including TMDCs, metal monochalcogenides, and h-BN, which can be grown by precise methods such as MBE [Dau18,Mor21,Cha20,Nak17,Lop21], we envision the realization of spintronic devices with dimensions and performance not achievable with conventional bulk materials.

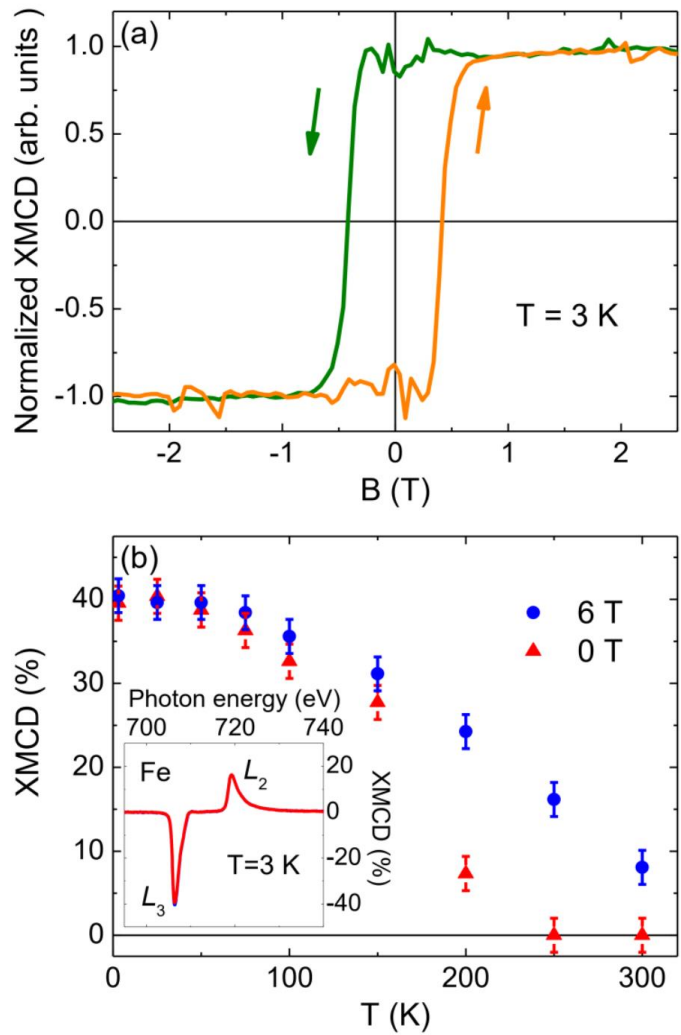

**Figure 4.** (a) Hysteresis loop recorded at the Fe $L_3$ edge for a Te-capped, ~20 nm thick FGT film on graphene/SiC(0001) at 3 K. The XMCD signal is normalized to its value in saturation. (b) Temperature dependence of the Fe $L_3$ XMCD maximum in saturation (6 T, blue circles) and remanence (red triangles). Inset: XMCD signal, expressed in percentage of the average XAS, recorded at 3 K under an out-of-plane magnetic field of 6 T (blue) and at remanence (red).

**Acknowledgments**

The authors would like to thank H.-P. Schönherr, C. Hermann, and C. Stemmler for their dedicated maintenance of the MBE system, as well as A. Riedel, W. Seidel, S, Rauwerdink, and A. Tahraoui for technical support. Furthermore, the authors appreciate the critical reading of the manuscript by Yukihiko Takagaki. They also acknowledge the provision of beamtime under the project HC-4068 at the European Synchrotron Radiation Facility (ESRF), located in Grenoble (France). ICN2 researchers acknowledge support from the European Union Horizon 2020 research and innovation programme under Grant Agreement No. 881603 (Graphene Flagship).

# *Supplementary Materials*

J. Marcelo J. Lopes,[1,‡] Dietmar Czubak,[1] Eugenio Zallo,[1,§] Adriana I. Figueroa,[2] Charles Guillemard,[3] Manuel Valvidares,[3] Juan Rubio Zuazo,[4,5] Jesús López-Sanchéz,[4,5] Sergio O. Valenzuela,[2,6] Michael Hanke,[1] Manfred Ramsteiner[1*]

1- Paul-Drude-Institut für Festkörperelektronik, Leibniz-Institut im Forschungsverbund Berlin e.V., Hausvogteiplatz 5-7, 10117 Berlin, Germany
2- Catalan Institute of Nanoscience and Nanotechnology (ICN2), CSIC and BIST, Campus UAB, Bellaterra, 08193 Barcelona, Spain
3- ALBA Synchrotron Light Source, Barcelona 08290, Spain
4- Spanish CRG BM25-SpLine at The ESRF – The European Synchrotron, 38000 Grenoble, France
5- Instituto de Ciencia de Materiales de Madrid (ICMM), CSIC, 28049 Madrid, Spain
6- Institució Catalana de Recerca i Estudis Avançats (ICREA), Barcelona 08010, Spain

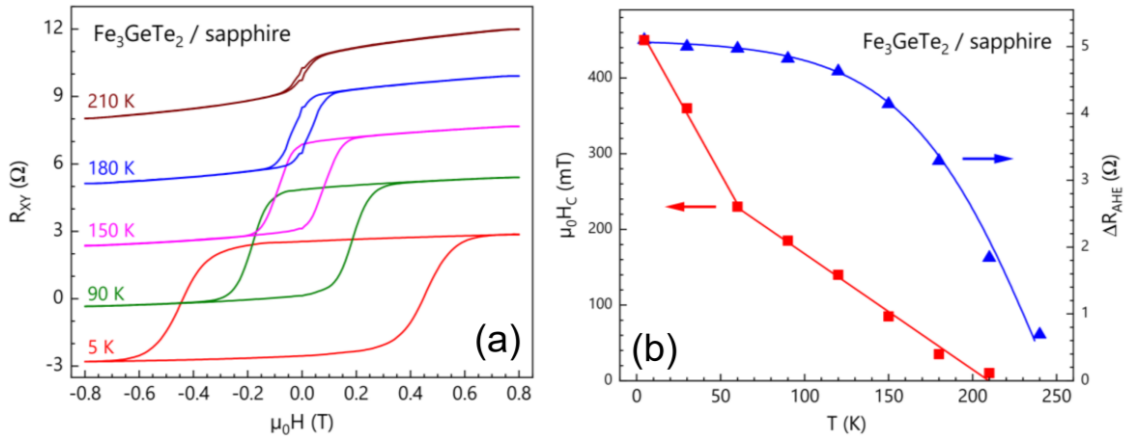

**Figure S1.** (a) Transverse resistance $R_{XY}$ of a 1 cm x 1 cm $Fe_3GeTe_2$/sapphire vdP structure during downward and upward magnetic field sweeps at different temperatures. The $R_{XY}$ curves for temperatures above 5 K are successively upshifted by 2.5 Ω for clarity. (b) Coercive field $H_C$ and saturation resistance $\Delta R_{AHE}$ extracted from the AHE contribution as a function of temperature. The solid lines are guides to the eye.

Note that, in contrast to $Fe_3GeTe_2$/graphene structures, the $Fe_3GeTe_2$ film constitutes the only transport channel when grown on sapphire substrates.

---

‡ Corresponding Authors: lopes@pdi-berlin.de; ramsteiner@pdi-berlin.de

§ Current affiliation: Walter Schottky Institut and Physik Department, Technische Universität München, Am Coulombwall 4, 85748 Garching, Germany